\documentclass[reprint,superscriptaddress,amsmath,amssymb,aps,prl]{revtex4-1}

\usepackage{graphicx}
\usepackage{graphics}
\usepackage{amsmath}
\usepackage{dcolumn}
\usepackage{bm}
\usepackage{subfig}

\begin{document}

\title{Quantum Floquet anomalous Hall states and quantized ratchet effect in one-dimensional dimer chain driven by two ac electric fields}

\author{Jin-Yu Zou}
\affiliation{Beijing National Laboratory for Condensed Matter Physics, Institute of Physics, Chinese Academy of Sciences, Beijing 100190, China.}
\author{Bang-Gui Liu}%
 \email{bgliu@iphy.ac.cn}
\affiliation{Beijing National Laboratory for Condensed Matter Physics, Institute of Physics, Chinese Academy of Sciences, Beijing 100190, China.}
\affiliation{School of Physical Sciences, University of Chinese Academy of Sciences, Beijing 100190, China}

\date{\today}

\begin{abstract}
In condensed matter physics, one of the major topics is to find out and classify interesting novel topological matters and phenomena. Topologically nontrivial systems can also be achieved by using periodical driving fields. On the other hand, ratchet effect can be used to collect useful work from fluctuations or realize directed transport in periodically-driven systems. Here, we promote a dimer chain by applying two mutually-perpendicular ac electric fields, and obtain an effective two-dimensional Hamiltonian in the low-frequency regime. We thereby derive quantum Floquet anomalous Hall (QFAH) conductance and then find a quantized ratchet effect in the resulting current along the chain. Our further analysis shows that these originate from the electric fields only. This lengthwise quantized ratchet effect without magnetic field should be useful to designing novel high-performance electronic applications.
\end{abstract}

\pacs{Valid PACS appear here}
\maketitle


Since the discovery of quantum Hall effect\cite{Klitzing1980}, finding topologically nontrivial matters and phenomena has become one of the major tasks for physicists\cite{Thouless1982, Haldane1988,Kane2005a,Kane2005,zhang1,zhang2,exp1,ti2d,fang1,fang2,ti3d-exp2,hasan-rmp,qi-rmp, rev2014, Bernevig2013,addrmp1,ahe1,qniu}. Unlike the traditional phase transition based on symmetry breaking and Landau-Ginzburg theory, the topological matter transforms from one phase to another with the change of its topological number, which can be winding number, Chern number, $Z_2$ number etc\cite{Avron1983, Kane2005a, Fu2006, Moore2007, Qi2008,hasan-rmp,qi-rmp}. One of the most exciting features of topologically nontrivial matters is the robust gapless edge states, which may be greatly useful in the electronic technology and quantum devices. Generally, given a topological material one can deduce the phase diagram by calculating the topological number in the presence of certain symmetry\cite{Schnyder2008,addrmp2}. While this procedure can be greatly enriched by periodically driving some materials with light or ac electric fields\cite{Kitagawa2010, Lindner2011, Gomez-Leon2013, Rudner2013, Gomez-Leon2014, Goldman2014, Sentef2015, foa,Quelle2016}. The time periodicity can provide the effective Hamiltonian an extra dimension and an effective electric field\cite{Gomez-Leon2013}.

Ratchets, defined as periodically driven spatially asymmetrical systems, can be used to collect useful work from fluctuations or harvest dissipated energy\cite{rat5,rat6}. On the other hand, spin and magnetic ratchets are used to produce directed spin and magnetic transports in the presence of magnetic field or spin magnetization\cite{rat1,rat2,rat4}. Ratchet effect is also found in graphene irradiated with ac electric field in the presence of an in-plane magnetic field, being able to produce a directed dc current\cite{rat3}. These effects are highly desirable for spintronic applications. Magnetic field or magnetization is necessary to producing such ratchet effects\cite{rat1,rat2,rat4,rat3}.

Here, we use a simple one-dimensional (1D) dimer chain driven by two ac electric fields to realize a Floquet effective two-dimensional (2D) Hamiltonian  and obtain quantum Floquet anomalous Hall (QFAH) conductance. This is in contrast to previous quantum Hall effects on 2D systems driven by periodical fields\cite{Kitagawa2011, Zhou2014, Dahlhaus2015}. Furthermore, we find a unidirectional ratchet current along the chain and show that its coefficient is quantized in units of the Hall conductance times the frequency of the main field. Our further analysis shows that this ratchet effect originates from the ac electric fields only. More detailed results will be presented in the following.

\begin{figure}
  \centering
  \includegraphics[clip, width=8cm]{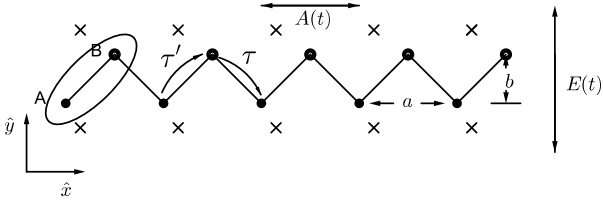}\\
  \caption{The zigzag dimer chain driven by two ac electric fields. The dimer chain is in the $\hat{x}$ direction and has $a$ as the lattice constance. There is a $\hat{y}$ coordinate difference $b$ between the A and B sites. $\tau$ and  $\tau'$ are the hopping constants. The $A(t)$ field is in the $\hat{x}$ direction and the $E(t)$ field in the $\hat{y}$ direction.}\label{model}
\end{figure}

Our model is a one-dimensional dimer chain driven by two mutually-perpendicular ac electric fields, as illustrated in Fig. \ref{model}. We define the $\hat{x}$ direction to be along the chain, and use $\tau$ and $\tau'$ to describe the two hopping parameters. One of the ac electric fields is described by a vector potential $A(t)=A_0\cos(\omega t)$ of frequency $\omega$ in the $\hat{x}$ direction, and the other is an ac electric field $E(t)=E_0\sin(2\omega t)$ in the $\hat{y}$ direction. The vector potential is equivalent to an electric field $A_0\omega\sin(\omega t)$. After making the standard Peierls substitution $\mathbf{k}\mapsto\mathbf{k}+e\mathbf{A}/\hbar$, the time-periodical Hamiltonian is given by
\begin{equation}\label{modelH}
  H(k,t)=(\tau'+\tau e^{-i(k+A(t))a})c_{kA}^\dagger c_{kB}+h.c.+c_{kA}^\dagger c_{kA}ebE(t)
\end{equation}
where the electron charge $e$ and Planck constant $\hbar$ are set to unit. Here, we suppose the dimer chain to be zigzag so that the A and B sites have different $\hat{y}$ coordinates, but the zigzag shape is not necessary for achieving nontrivial properties. It should be pointed out that the ac electric field $E(t)$ is necessary to make the model become topologically nontrivial. Without the $E(t)$ field, in high frequency regime ($\omega\gg\tau,\tau'$) the time-dependent model is equivalent to a time-independent dimer chain with hopping parameters renormalized by the amplitude of $A(t)$, and in low frequency regime ($\omega\ll\tau,\tau'$) the system can be treated as a 2D lattice belonging to BDI class and is trivial \cite{Gomez-Leon2013}. In contrast, the $E(t)$ field causes different electric potentials between A and B sites, and thereby can make the system become topologically nontrivial\cite{Schnyder2008}.

We shall investigate the model (1) by using the Floquet theory and the approach developed by G\'{o}mez-Le\'{o}n and Platero \cite{Gomez-Leon2013}.
With the Floquet-Bloch ansatz\cite{Platero2004}, the states of a periodically driven Hamiltonian $H(x,t+T)=H(x,t)$ can be expressed as $|\Psi_{\alpha,k}(x,t)\rangle=e^{ik\cdot x-i\epsilon_{\alpha,k}t}|u_{\alpha,k}(x,t)\rangle$\cite{Gomez-Leon2013,zjy}, with the band index $\alpha$, the quasienergy $\epsilon_{\alpha,k}$, and the periodical Floquet-Bloch states $|u_{\alpha,k}(x,t+T)\rangle=|u_{\alpha,k}(x,t)\rangle$. Thus the time-dependent Schr\"{o}dinger equation reduces to
\begin{equation}\label{HF}
  [H(k,t)-i\partial_t]|u_{\alpha,k}\rangle=\epsilon_{\alpha,k}|u_{\alpha,k}\rangle
\end{equation}

We can define $H_F=H(k,t)-i\partial_t$ as the Floquet operator, and expand the states in the composed Hilbert space $\mathcal{S}=\mathcal{H}\otimes\mathcal{T}$\cite{Sambe1973}, where $\mathcal{H}$ is the ordinary Hilbert space and $\mathcal{T}$ is the T-periodical function space coming form the Fourier transformation $|u_{\alpha,k}(t)\rangle=\sum_n|u_{\alpha,k,n}\rangle e^{in\omega t}$ and $c_{\alpha,k}(t)=\sum_n c_{\alpha,k,n}e^{in\omega t}$. With the inner product $\langle\langle\cdot\cdot\cdot\rangle\rangle=\frac{1}{T}\int^T_0\langle\cdot\cdot\cdot\rangle dt$, $H_F$ can be expressed in space $\mathcal{S}$:
\begin{equation}\label{HFnm}
  \langle n|H_F|m\rangle=\frac{1}{T}\int^T_0H(k,t)e^{-i(n-m)\omega t}dt - n\omega \delta_{n,m}
\end{equation}
Consequently, the matrix elements of the Floquet operator can be written as
\begin{equation}\label{effectiveH}
\begin{split}
(\mathcal{H}_F)_{n,m} = & B_{nm}(k_x)c_{k_xAn}^\dagger c_{k_xBm}+B^\ast_{nm}(k_x)c_{k_xBn}^\dagger c_{k_xAm} \\
                       &+(\delta_{n,m+2}-\delta_{n,m-2})iDc_{k_xAn}^\dagger c_{k_xAm} \\
&-n\omega \delta_{n,m}(c_{k_xA n}^\dagger c_{k_xA m}+c_{k_xB n}^\dagger c_{k_xB m})
\end{split}
\end{equation}
where we have $D=ebE_0\pi/\omega$ and $B_{nm}=\tau'[\delta_{n,m}+\lambda e^{-ik_xa}J_{n-m}(A_0a)e^{-i\frac{\pi}{2}(n-m)}]$, with $\lambda=\tau/\tau'$, and $J_l(\alpha)$ is the $l$th Bessel function of the first kind. For simplicity, we use $J_l$ to denote $J_l(A_0a)$.

\begin{figure}
  \centering
  \includegraphics[clip, width=7.5cm]{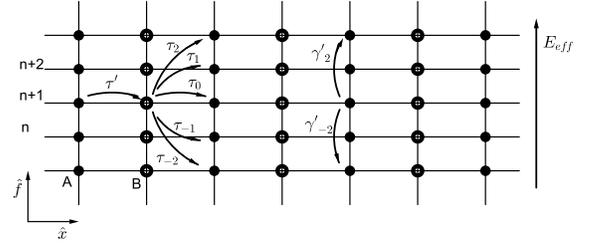}\\
  \caption{The effective 2D lattice of the periodically driven dimer chain. The additional $\hat{f}$  dimension is spanned by $\{ n\}$ due to the periodicity. Other hopping constants of higher orders are neglected because of $A_0a \le 1$. The effective electric field is $E_{\rm eff}=\omega $.}\label{el}
\end{figure}

To investigate the Hamiltonian Eq. (\ref{effectiveH}), one can take $\{n\}$ to form an additional dimension in the $\hat{f}$ direction and thus treat the Hamiltonian as an effective 2D TB model in the 'real' space ($\hat{x}$,$\hat{f}$), as illustrated in Fig. \ref{el}. Accordingly, we have effective hopping parameters: $\tau_0=\tau J_0$, $\tau_{\pm1}=-i\tau J_1$, $\tau_{\pm2}=-\tau J_2$, and $\gamma'_{\pm2}=\pm iD$. Usually, we restrict the strength of field by requiring $A_0a \le 1$ to make sure that the $J_3$ terms and so on are small enough to be neglected. The last term of Eq. (\ref{effectiveH}) will be equivalent to an effective electric field $E_{\rm eff}=\omega $ in the $\hat{f}$ direction.
Transforming the frequency space $\{m\}$ in the $\hat{f}$ direction to the corresponding 'momentum' space $k_f$ (taking values in $[-\pi,\pi]$), we obtain an effective 2D $k$ space, ($k_x,k_f$). It can be proved that the Hamiltonian is diagonal in $k_x$. Because we focus on low frequency dynamics, the last term in Eq. (4) can be treated as a perturbation, and will be taken into account later. Defining $c_k=(c_{kA}, c_{kB})^T$, we can express the remainder of Eq. (4) as $c_k^\dagger H_{0}(k)c_k$. Thus, the effective matrix Hamiltonian $H_{0}(k)$ is given by
\begin{equation}\label{Heff}
  H_{0}(k_x,k_f)=d_x\sigma_x+d_y\sigma_y+d_z\sigma_z,
\end{equation}
where the coefficients are defined as: $d_x=\tau^\prime+\tau[J_0-2J_2\cos(2k_f)]\cos(k_xa)-2\tau J_1\sin(k_xa)\cos{k_f}$,
$d_y=\tau[J_0-2J_2\cos(2k_f)]\sin(k_xa)+2\tau J_1\cos(k_xa)\cos{k_f}$, and
$d_z=-D\sin(2k_f)$.

\begin{figure}
  \centering
  \includegraphics[clip, width=7cm]{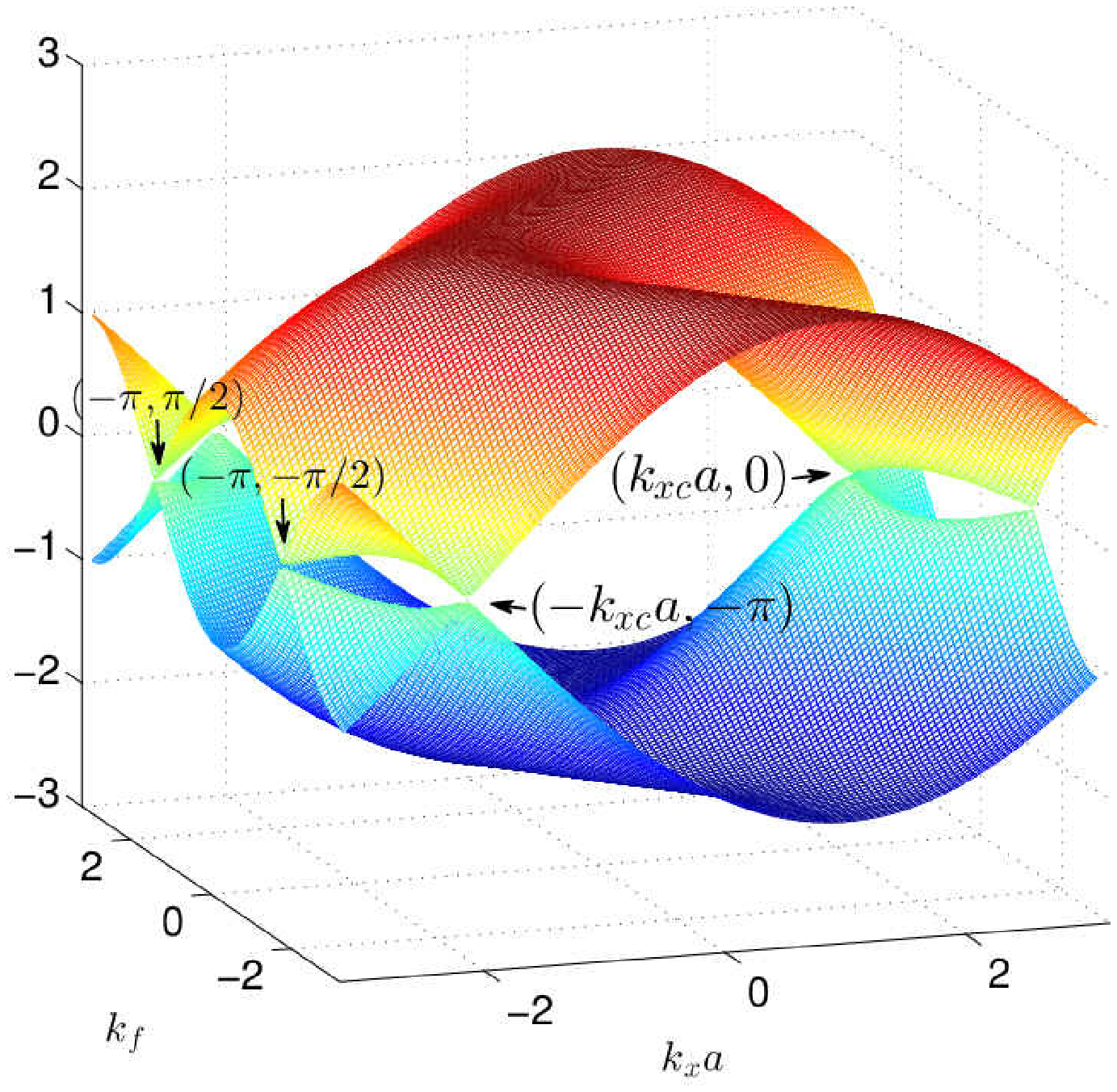}\\
  \caption{The band structure of the effective Hamiltonian, with the four Dirac points labelled. The parameters are set to $\tau=\tau^\prime=1$, $A_0a=1$, and $D=0.2$.}\label{band}
\end{figure}

Interestingly, the effective Hamiltonian represents a torus in the ($d_{x}$,$d_{y}$,$d_{z}$) space. It should be noted that this torus is a double torus, i.e. each point on the torus maps to two points in the first Brillouin zone of the momentum space ($k_x,k_f$). Therefore, the Hamiltonian has four extremum points: $\mathbf{K}=(\pi/a,-\pi/2)$, $\mathbf{K}'=(\pi/a,\pi/2)$, $\mathbf{\Gamma}=(k_{xc},0)$, and $\mathbf{\Gamma}'=(-k_{xc},\pi)$.
$\mathbf{K}$ and $\mathbf{K}'$ map to the point ($1-\lambda/W_1$,0,0) in the ($d_{x}$,$d_{y}$,$d_{z}$) space, and $\mathbf{\Gamma}$ and $\mathbf{\Gamma}'$ to another point ($1-\lambda/W_2$,0,0), where $W_1$ and $W_2$ are defined by $W_1=1/(J_0+2J_2)$ and $W_2=1/\sqrt{(2J_1)^2+(J_0-2J_2)^2}$. The value of $k_{xc}$ will be given in the following. Near each of the four points, the effective low-energy physics can be described with neither Dirac nor Weyl equation, but the first (last) two points are degenerate in energy. The local properties near $\mathbf{K}$ and $\mathbf{K}'$ can be described with four-component states including valley pseudo-spin freedom, and so can $\mathbf{\Gamma}$ and $\mathbf{\Gamma}'$. In addition, they are gapped unless the zero mass condition is satisfied. Therefore, we can denote them by Dirac points for convenience. Because of $W_1\ne W_2$ for $A_0a>0$, we can have two massless Dirac points at most for a $\lambda$ value. For $W_1\ne \lambda \ne W_2$, the quasi-energy bands are gapped. We illustrate the band structure and label the four Dirac points in Fig. \ref{band}.

Following the standard approach to quantum Hall effect, we can derive a Hall conductance for the ($\hat{x}$,$\hat{f}$) space through the Chern number $N_C$ by integrating the curvature $F(k)$ of Berry phase $A_j(k)$ over the 2D Brillouin zone,
\begin{equation}\label{chern}
\left\{\begin{split}
 & \sigma_{\rm 2D}=\frac{e^2}{h}N_C=\frac{e^2}{2\pi h}\int dk_xdk_f F(k)\\
 & F(k)=\frac{\partial A_f(k)}{\partial k_x}-\frac{\partial A_x(k)}{\partial k_f}\\
 & A_j(k)=-i\langle\psi_g(k)|\frac{\partial}{\partial k_j}|\psi_g(k)\rangle
\end{split}\right.
\end{equation}
where $j$ can take $\hat{x}$ or $\hat{f}$. The ground-state wavefunction of $H_{0}$, $|\psi_g(k)\rangle=(\psi_{gA}(k),\psi_{gB}(k))^T$, is obtained from the Hamiltonian (5). One should be careful because the $U(1)$ gauge of the states is not always well defined on the whole BZ unless the Chern number is zero. One should cautiously divide the BZ into different regions and set down the gauge separately.

Alternatively, we can calculate the Hall conductance by using a physically transparent method\cite{Bernevig2013}. With the variation of parameter $\lambda$, the torus will move along the $d_x$ axis. When the origin point is outside the torus, the Hamiltonian is trivial. When the origin point penetrates the torus, the mass term will change sign at the massless Dirac points. As a result, in order to study the topological property of the torus, we need only to investigate the helicity of the massless Dirac points. For brevity, $\tau^\prime$ will be set to 1 in the following. When the condition of $\lambda=W_1$ is satisfied, $\mathbf{K}$ and $\mathbf{K}'$ become massless Dirac points.
Expanded near one of the two massless Dirac points, the Hamiltonian (5) can be expressed as
\begin{equation}\label{Dirac}
  H(K_D) = m_K\sigma_x -\lambda(\frac{k_x}{W_1}+2\epsilon_D J_1k_f)\sigma_y+2Dk_f\sigma_z
\end{equation}
where $m_K=(1-\frac{\lambda}{W_1})$, $K_D$ is either $\mathbf{K}$ or $\mathbf{K}'$, and $\epsilon_D$ is equivalent to +1 or -1, accordingly. The $k$ variable in Eq. (7) is defined with respect to $K_D$.
By rotating anticlockwise the local Hamiltonian along the $y$ axis, it can be rewritten as the standard form, $H(K_D)=\sum_{i,j}A_{i,j}(K_D)k_i\sigma_j+m_K\sigma_z$, where the sum is over $x$ and $y$. The resulting matrix $A$ can be given by
\begin{equation}\label{rotated}
 A(K_D)=\left(
      \begin{array}{cc}
        0 & -\frac{\lambda}{W_1} \\
        -2D & -2\lambda \epsilon_D J_1
      \end{array}
    \right)
\end{equation}
Except the different signs for $\epsilon_D$, the local Hamiltonian near the two massless Dirac pints have the same form.
It can be proved that the two massless Dirac points have the same helicity: $\textit{sgn}(\det[A_{ij}])=-1$. As is well known\cite{Bernevig2013}, when the mass term of a Dirac point changes from $M_0$ to $M_1$, the Hall conductance will have a change: $\Delta\sigma_{xy}=\frac{e^2}{2h}\textit{sgn}(\det[A_{ij}])[\textit{sgn}(M_1)-\textit{sgn}(M_0)]$. As a result, when the parameter $\lambda-W_1$ changes from negative to positive values, the Hall conductance of the effective 2D lattice model will increase by $2e^2/h$.

When $\lambda=W_2$ is satisfied, $\mathbf{\Gamma}$ and $\mathbf{\Gamma}'$ become massless Dirac points. The same discussion applies to these two massless Dirac points. From Eq. (\ref{Heff}), we obtain $\tan{k_{xc}}=-2J_1\epsilon^\prime_D/(J_0-2J_2)<0$, where $\epsilon^\prime_D$ is equivalent to 1 and -1, respectively. $k_{xc}$ lies in the second quadrant for $\epsilon^\prime_D=1$, or in the third quadrant for $\epsilon^\prime_D=-1$. The same effective 2D Hamiltonians near both $\mathbf{\Gamma}$ and $\mathbf{\Gamma}'$ are derived:
\begin{equation}\label{Dirac2}
  H(\Gamma_D) = m_{\Gamma}\sigma_x
  + \frac{\lambda}{W_2}k_x\sigma_y -2Dk_f\sigma_z,
\end{equation}
where $m_{\Gamma}=1-\frac{\lambda}{W_2}$. The standard form of the $A$ matrix reads
\begin{equation}\label{rotated2}
  A(\Gamma_D)=\left(
      \begin{array}{cc}
        0 & -\frac{\lambda}{W_2} \\
        2D & 0
      \end{array}
    \right)
\end{equation}
Consequently, when the parameter $\lambda-W_2$ changes from negative to positive value, the Hall conductance of the effective 2D lattice model decreases by $2e^2/h$.

\begin{figure}
  \centering
  \includegraphics[clip, width=7cm]{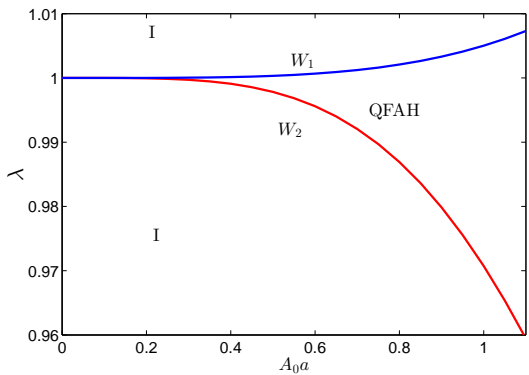}\\
  \caption{The ($A_0a$,$\lambda$) phase diagram. The quantum Floquet anomalous Hall (QFAH) phase is determined by $W_2<\lambda<W_1$, and elsewhere is the trivial insulating (I) phase.}\label{tau}
\end{figure}

When $\lambda$ is equivalent to $0$ or $\infty$, the effective 2D lattice is definitely trivial. The analysis above implies that when $\lambda$ is between $W_1$ and $W_2$, i.e. $W_2<\lambda<W_1$, the effective 2D $\hat{x}$-$\hat{f}$ lattice model has a QFAH phase with Hall conductance $\sigma_{\rm 2D}=-2e^2/h$; and the Hall conductance is quantized in units of $2e^2/h$, with the Chern number being -1. For $\lambda>W_1$ or $\lambda<W_2$, the effective 2D lattice is trivial and causes zero Hall conductance, with the Chern number being 0. We present the topological phase diagram in Fig. 4. Because this quantized Hall conductance is derived in the absence of any magnetic field, these are actually quantum Floquet anomalous Hall QFAH states for the ($\hat{x}$,$\hat{f}$) space, as is usually  defined \cite{ahe1,qniu}.


Now we address the last term in Hamiltonian (4). It can be expressed as $c_k^\dagger H^{\prime}(k)c_k$, and can be proved\cite{Gomez-Leon2013} to be equivalent to an effective electric field $E_{\rm eff}=\omega$ in the additional $\hat{f}$ dimension. Because we already have the quantized anomalous Hall conductance $\sigma_{\rm 2D}=-2e^2/h$ in the effective 2D $\hat{x}$-$\hat{f}$ space, we can naturally obtain a unidirectional current $j_x(n)=j_x=\sigma_{\rm 2D}\omega$ in the $\hat{x}$ direction. It is interesting that $j_x(n)$ is a constant, independent of both of the coordinates in the 2D $\hat{x}$-$\hat{f}$ space. The Fourier transformation from the composed Hilbert space to the real space leads to the  time-dependent one-dimensional current:
\begin{equation}\label{j(t)}
  j_x(t)=\sum_nj_x(n)e^{-in\omega t}=j_x\sum_m\delta(t-mT)
\end{equation}
Because of $j_x<0$, we can conclude that the current is unidirectional in the $\hat{x}$ dimension and periodical in time $t$, as demonstrated in Fig. 5. The average value of the current over one period is equivalent to $j_x=\sigma_{\rm 2D}\omega$. Considering that our dimer chain can have the inversion symmetry in the $\hat{x}$ axis for $\lambda=1$ and the two driving fields are purely ac electric fields, this periodical unidirectional current makes a lengthwise ratchet current. Because the Floquet anomalous Hall conductance is quantized, $j_x$ is also quantized in units of $\sigma_{\rm 2D}\omega$. Therefore, this is a quantized ratchet effect in the $\hat{x}$ axis.

\begin{figure}
  \centering
  \includegraphics[clip, width=7cm]{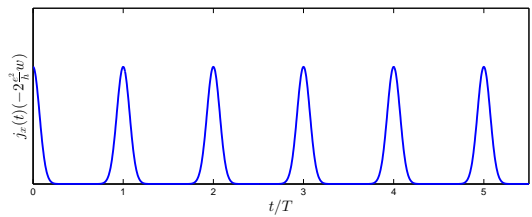}\\
  \caption{The time dependence of the lengthwise ratchet current for the nontrivial phase defined by $W_2<\lambda<W_1$ in the phase diagram.}\label{tau}
\end{figure}

For open boundary condition for the dimer chain, the effective 2D model will possess edge states if it is topologically nontrivial. Considering that the model is still periodical in the $\hat{f}$ direction, the condition of existence of edge states is the nonzero winding numbers of the effective Hamiltonian which is the function of $k_f$ \cite{Gomez-Leon2013}.


The second ac electronic field $E(t)=E_0\sin(2\omega t)$, causing the $D$ term in $H_{0}$ of Eq.(\ref{Heff}), breaks the time reversal symmetry of the effective 2D lattice, i.e. $H^{*}_{0}(k)\neq H_{0}(-k)$. Without the $D$ term, we should have $\gamma'_{\pm2}=0$ and the electron hopping along a circle does not get a nonzero phase, but with the $D$ term added, the electron hopping through $\tau_2$, $\gamma'_{-2}$ and $\tau^{\dagger}_0$ will get a nonzero phase $\pi/2$. Thus the phase factor on the hopping parameters cannot be removed by gauge transformation, which implies that the electron hopping breaks time reversal symmetry. It is necessary to make the frequency of $E(t)$ twice that of $A(t)$. If $E(t)$ would have the frequency of $\omega$, the symmetry $H_{0}(k_x,k_f)=H_{0}(k_x,\pi-k_f)$ should make the Hamiltonian a cylinder, which is trivial all the time. The same reason holds for other odd numbers times $\omega$. Even numbers times $\omega$ for $E(t)$ can make nontrivial torus Hamiltonian, but the double frequency is the best for simplicity.

Our 1D dimer model can be easily realized with Polyacetylene [(CH)$_x$], used in original Su-Schrieffer-Heeger model\cite{Su1979,add1}. It is reasonable to use typical parameters: $\lambda\simeq 1$, $\tau\simeq 2.2$eV, and $a\simeq 2.4$\AA. In addition, it can be realized in some 1D adatom systems on semiconductor surfaces\cite{add2} and some cold atom systems. The condition of strength for the $A(t)$ field, $A_0a\le 1$, can be easily satisfied, and especially when the parameter $\lambda$ is near 1, $A_0a\ll 1$ can be realized. It is certain that the condition of low frequency, $\omega\ll \tau,\tau^\prime$, can be satisfied. On the other hand, because the ratchet current is proportional to the frequency $\omega$, it is reasonable to take as large $\omega$ as possible. As for the strength of the $E(t)$ field, any nonzero value is enough to achieve the QFAH states and the quantized ratchet effect.

In summary, we have promoted a 1D dimer chain by applying two mutually-perpendicular ac electric fields and obtained an effective 2D Hamiltonian (with a time-related extra dimension) in the low frequency regime. We have shown that the 2D effective model hosts a QFAH phase featuring a quantized Hall conductance and the QFAH states cause a unidirectional ratchet current along the chain. The lengthwise current is quantized in units of the Hall conductance times the frequency of the field $A(t)$. Our further analysis indicates that because no magnetic field is applied and the inversion symmetry in the $\hat{x}$ axis is kept for $\lambda=1$, the QFAH states and the quantized ratchet effect originate from the electric fields only.
The lengthwise quantized ratchet effect without magnetic field can be used to design novel high-performance electronic applications.

\begin{acknowledgments}
This work is supported by the Nature Science Foundation of China (Grant No. 11574366), by the Strategic Priority Research Program of the Chinese Academy of Sciences (Grant No.XDB07000000), and by the Department of Science and Technology of China (Grant No. 2016YFA0300701).
\end{acknowledgments}


\end{document}